\documentstyle[12pt]{article}

\catcode`\@=11
\long\def\@makefntext#1{
\protect\noindent \hbox to 3.2pt {\hskip-.9pt
$^{{\ninerm\@thefnmark}}$\hfil}#1\hfill}		

 \def\@makefnmark{\hbox to 0pt{$^{\@thefnmark}$\hss}}  

\def\ps@myheadings{\let\@mkboth\@gobbletwo
\def\@oddhead{\hbox{}
\rightmark\hfil\ninerm\thepage}
\def\@oddfoot{}\def\@evenhead{\ninerm\thepage\hfil
\leftmark\hbox{}}\def\@evenfoot{}
\def\sectionmark##1{}\def\subsectionmark##1{}}

\newcounter{sectionc}\newcounter{subsectionc}\newcounter{subsubsectionc}
\renewcommand{\section}[1] {\vspace{0.6cm}\addtocounter{sectionc}{1}
\setcounter{subsectionc}{0}\setcounter{subsubsectionc}{0}\noindent
	{\bf\thesectionc. #1}\par\vspace{0.4cm}}
\renewcommand{\subsection}[1] {\vspace{0.6cm}\addtocounter{subsectionc}{1}
	\setcounter{subsubsectionc}{0}\noindent
	{\it\thesectionc.\thesubsectionc. #1}\par\vspace{0.4cm}}
\renewcommand{\subsubsection}[1]
{\vspace{0.6cm}\addtocounter{subsubsectionc}{1}
	\noindent {\rm\thesectionc.\thesubsectionc.\thesubsubsectionc.
	#1}\par\vspace{0.4cm}}

\newcounter{appendixc}
\newcounter{subappendixc}[appendixc]
\newcounter{subsubappendixc}[subappendixc]

\renewcommand{\appendix}[1] {\vspace{0.6cm}
        \refstepcounter{appendixc}
        \setcounter{figure}{0}
        \setcounter{table}{0}
        \setcounter{equation}{0}
        \renewcommand{\thefigure}{\Alph{appendixc}.\arabic{figure}}
        \renewcommand{\thetable}{\Alph{appendixc}.\arabic{table}}
        \renewcommand{\theappendixc}{\Alph{appendixc}}
        \renewcommand{\theequation}{\Alph{appendixc}.\arabic{equation}}
        \noindent{\bf Appendix \theappendixc #1}\par\vspace{0.4cm}}

\def\abstracts#1{{
	\centering{\begin{minipage}{30pc}\tenrm\baselineskip=12pt\noindent
	\centerline{\tenrm ABSTRACT}\vspace{0.3cm}
	\parindent=0pt #1
	\end{minipage}}\par}}

\renewenvironment{thebibliography}[1]
	{\begin{list}{\arabic{enumi}.}
	{\usecounter{enumi}\setlength{\parsep}{0pt}
\setlength{\leftmargin 1.25cm}{\rightmargin 0pt}
	 \setlength{\itemsep}{0pt} \settowidth
	{\labelwidth}{#1.}\sloppy}}{\end{list}}

\newcounter{itemlistc}
\newcounter{romanlistc}
\newcounter{alphlistc}
\newcounter{arabiclistc}

\newcommand{\fcaption}[1]{
        \refstepcounter{figure}
        \setbox\@tempboxa = \hbox{\tenrm Fig.~\thefigure. #1}
        \ifdim \wd\@tempboxa > 6in
           {\begin{center}
        \parbox{6in}{\tenrm\baselineskip=12pt Fig.~\thefigure. #1}
            \end{center}}
        \else
             {\begin{center}
             {\tenrm Fig.~\thefigure. #1}
              \end{center}}
        \fi}

\newcommand{\tcaption}[1]{
        \refstepcounter{table}
        \setbox\@tempboxa = \hbox{\tenrm Table~\thetable. #1}
        \ifdim \wd\@tempboxa > 6in
           {\begin{center}
        \parbox{6in}{\tenrm\baselineskip=12pt Table~\thetable. #1}
            \end{center}}
        \else
             {\begin{center}
             {\tenrm Table~\thetable. #1}
              \end{center}}
        \fi}

\def\@citex[#1]#2{\if@filesw\immediate\write\@auxout
	{\string\citation{#2}}\fi
\def\@citea{}\@cite{\@for\@citeb:=#2\do
	{\@citea\def\@citea{,}\@ifundefined
	{b@\@citeb}{{\bf ?}\@warning
	{Citation `\@citeb' on page \thepage \space undefined}}
	{\csname b@\@citeb\endcsname}}}{#1}}

\newif\if@cghi
\def\cite{\@cghitrue\@ifnextchar [{\@tempswatrue
	\@citex}{\@tempswafalse\@citex[]}}
\def\citelow{\@cghifalse\@ifnextchar [{\@tempswatrue
	\@citex}{\@tempswafalse\@citex[]}}
\def\@cite#1#2{{$\null^{#1}$\if@tempswa\typeout
	{IJCGA warning: optional citation argument
	ignored: `#2'} \fi}}

\def\fnt#1#2{\footnotetext{\kern-.3em
	{$^{\mbox{\sevenrm #1}}$}{#2}}}

 1
 1
 1

\font\tenbf=cmbx10
\font\tenrm=cmr10
\font\tenit=cmti10

\font\ninerm=cmr9

\let\sll=\l			

\def\d{\delta}

\def\j{\psi}
\def\l{\lambda}
\def\m{\mu}

\def\J{\Psi}
\def\L{\Lambda}

\def\sfA{{\sf A}}                       \def\sfa{{\sf a}}
\def\sfB{{\sf B}}			\def\sfb{{\sf b}}

\def\sfH{{\sf H}}

			\def\sfp{{\sf p}}

\def\cd{{\cal D}}

\def\cg{{\cal G}}
\def\ch{{\cal H}}

\def\ck{{\cal K}}

\def\car{{\cal R}}

\def\bd{\begin{displaymath}}
\def\ed{\end{displaymath}}
\def\be{\begin{equation}}
\def\ee{\end{equation}}
\def\bq{\begin{eqnarray}}
\def\eq{\end{eqnarray}}
\def\ba{\begin{array}}
\def\ea{\end{array}}
\def\bqn{\begin{eqnarray*}}
\def\eqn{\end{eqnarray*}}
\def\half{\frac{1}{2}}

\def\lbl{\label}
\def\dd{\partial}
\def\rar{\rightarrow}

\def\half{\frac12}

\newcommand{\lp}{\left(}
\newcommand{\rp}{\right)}

\newcommand{\bi}[1]{\bibitem{#1}}

\newcommand{\der}[1]{\frac{\partial}{\partial #1}}
\newcommand{\dr}[2]{\frac{\partial #2}{\partial #1}}
\newcommand{\vder}[2]{\frac{\d #1}{\d h_{#2}}}

\begin{document}

\centerline{\tenbf QUANTUM POTENTIAL AND QUANTUM GRAVITY}
\vspace{0.8cm}
\centerline{\tenrm J. KOWALSKI--GLIKMAN}
\baselineskip=13pt
\centerline{\tenit Institute for Theoretical Physics, University of
Wroc\sll{}aw}
\baselineskip=12pt
\centerline{\tenit Pl. Maxa Born 9, 50204 Wroc\sll{}aw, Poland\footnote{%
e-mail address:
jurekk@ift.uni.wroc.pl and jurekk@fuw.edu.p}}
\vspace{0.9cm}
\abstracts{
The quantum potential approach makes it possible to construct a
complementary picture of quantum mechanical evolution which reminds
classical equation of motion. The only difference as compared to
equations of motion for the underlying classical system is the presence
of an additional potential term being a functional of the real part of
the wavefunction. In the present paper this approach is applied to the
quantum theory of gravity based on Wheeler -- De Witt equation. We describe
the derivation of the `quantum Einstein equation' and discuss the new
features of their solutions.}

\section{Preface}

It gives me great pleasure to contribute this article to the volume in
honor of Professor Jerzy Lukierski. I clearly remember when,
still being a student, I have met him for the first time
during one of the winter Karpacz schools in late 1970s. Since
then he was always close to me, first as an advisor
and referee of my PhD thesis, and then as a close collaborator in many
research projects and a very good friend. I cannot possibly fully pay my
debt to him; let this article be at least an expression of
acknowledgement of what I owe him.

\section{Introduction}

The enigma of quantum gravity is probably the most challenging problem
of modern theoretical physics (for the recent reviews see \cite{Isham1},
\cite{Isham2}, \cite{Isham3}, and \cite{Ashtekar}). It is for the difficulty
of the problem
that in spite of its importance, a very little progress has been made
so far. Among many approaches, the canonical quantization seems to be
the most natural one for at least  two following reasons. First
of all, we understand quite
well the
point we start at: the Einstein theory of general relativity and the Dirac
theory of quantization of constrained systems. Secondly, other more
exotic approaches (most notably the string theory)  introduce, as a rule,
a number of their own difficulties.

It is the lesson resulting from the modern developments that
the quantum theory of gravity must be non-perturbative. This fact comes
from the careful analysis of the condition of diffeomorphism invariance.
{}From this point of view the non-renormalizibility of perturbative
quantum gravity is not surprising; indeed the modern interpretation of
this fact is that it just shows that the perturbative expansion
of quantum gravity does not make sense.

Most of the recent work on the canonical approach to quantum gravity is
related to the loop formulation (see \cite{Ashtekar} and \cite{Rovelli}.)
This formulation has the virtue that
some part of constraints (Gauss law) is in terms of loop variables
solved automatically. However, it turned out that this formulation
has its own problems, the major of whose are the following:
\begin{enumerate}
\item In quantum theory all the composite operators require
regularizations. In particular it turns out that in the loop representation
the metric is a
composite operator and there seems to be no regularization which is consistent
with diffeomorphism invariance of the theory. Therefore, even if the
theory can be eventually quantized in this representation, it is
presumably not equivalent to the quantization in metric representation.
(One may argue however that the tetrad representation, which is the starting
point for Ashtekar approach, is the more
fundamental because only in this representation the coupling to fermions
exists).
\item It is well known that contrary to the quantum mechanical case,
there exist many representations of quantum field theory which are not
equivalent. Thus one must ask the question as to if some
prediction of the theory like quantization of volume and area operators
\cite{rovsmo},
which are claimed to indicate the discrete structure of space-time at
sub-Planckian regime, are solid facts and not artifacts of the chosen
representation. As a matter of fact, it is unclear if in this formalism
one can construct any operator with continuous spectrum at all.
\item One of the major arguments in favor of using Ashtekar variables
was the simplicity of the hamiltonian constraints. But it turned out
that because of the absence of background metric it is very hard to
regularize this operator, and solving the regularized version of it
may well be as
complicated as it is in the case of the Wheeler-De Witt operator.
\end{enumerate}

I made the comments above just to justify the statement that it is not
outrageous to be a little bit old fashioned and base our discussion on
the De Witt formulation of quantum gravity \cite{BDW}. For, if both
metric and Ashtekar formulations are equivalent, using them we look at
physical reality from two equivalent and yet distinct points of view.
If, however, these formulations are not equivalent, they will compete as
to which is the correct one. In both cases, therefore, investigating the
prediction of metric formulation is quite important. I would like to
stress however that the results described below may well be applied to
some other formulations of quantum gravity.

Quantum gravity theory faces yet another set of problems. Assume that
sooner or later we will have in our possession a class of solutions of
this theory. These solutions will be presumably of the form of some
wavefunction being a solution of a set of complicated equations defining
the theory. Now the question is how should we interpret such a
wavefunction? It seems that if we accept the orthodox interpretation of
quantum mechanics, we will immediately face a number of problems like:
what is the meaning of the wavefunction, how are we to interpret
superpositions of states (universes) etc? There are many attempts to
address these kind of problems (see e.g., \cite{Isham2}).

In my personal opinion it would be fruitful to try to translate the
information carried by the wavefunction to the language which would be
easier to comprehand. It happens that such a language exists and is
provided by the 'pilot wave' or 'quantum potential' approach to quantum
mechanics (see \cite{Bohm},\cite{Bell}, \cite{four}). Before discussing
the merit of this approach as applied to quantum gravity, to set the
stage, let us consider some simple examples.
\clearpage

\section{Quantum potential in action: quantum mechanical examples}
\subsection{The standard quantum mechanical system}

Let us consider the simple one dimensional quantum mechanics. The time
evolution
of the wave function of particle of mass $m$ is governed by the Schr\"odinger
equation
\be
i\hbar\der{t}\J = -\frac{\hbar^2}{2m}\frac{\dd^2}{\dd x^2}\J + V(x)\J.
\ee
Let us now consider the polar decomposition of the wave function
$$
\J(x,t) = R(x,t)\exp\lp\frac{i}{\hbar}S(x,t)\rp,
$$
where both $R$ and $S$ are real functions. Substituting this into the
Schr\"odinger
equation, we obtain two equations for real and imaginary parts, to wit
\be
\der{t}S+\frac{1}{2m}\lp\frac{\dd S}{\dd x}\rp^2 + V(x) -
\frac{\hbar^2}{2m}\frac{1}{R}\frac{\dd^2 R}{\dd x^2}=0
\lbl{real}
\ee
and
\be
\der{t}R+\frac{1}{m}\frac{\dd R}{\dd x}\frac{\dd S}{\dd x}
+\frac{1}{2m}R\frac{\dd^2 S}{\dd x^2}=0\lbl{im}
\ee
The interpretation of Eq.\ (\ref{im}) is clear: this is noting but the
continuity equation. Eq.\ (\ref{real}), on the other hand, can be interpreted
as the Hamilton-Jacobi equation, with the additional term proportional to
$\hbar^2$ and resulting from the wave function of the system. This potential
is called the `quantum potential' and it is customary to denote it by $Q(x,t)$.
Observe that this potential is, in general, time dependent. It is a well known
fact that the Hamilton-Jacobi equation contains the whole dynamics of the
system.
Thus, interpreting Eq.\ (\ref{real}) in such a way, we can immediately write
Hamilton equations of motion:
\be
\dot p =\left\{ p, \ch\right\},\lbl{h1}
\ee
\be
\dot x =\left\{ x, \ch\right\},\lbl{h2}
\ee
where
\be
\ch =
\frac{p^2}{2m} + V(x) -
\frac{\hbar^2}{2m}\frac{1}{R}\frac{\dd^2 R}{\dd x^2}
\ee
is the effective hamiltonian containing the quantum potential term.

Let us discuss interpretation of these equations. To set the stage, let us
consider Schr\"odinger equation first. Here we have to provide one
initial condition for $\J$, $\J_0=\J(x,t=0)$ which corresponds to two
initial conditions for $R$ and $S$. It should be observed that Eq.\%
(\ref{im}) expresses conservation of probability.

Now, in the quantum potential approach the situation is quite different.
First of all, Hamilton's equations (\ref{h1}), (\ref{h2}) govern the
evolution of $x(t)$, $p(t)$. We need therefore two initial conditions
for $p$ and $x$. Second, we have Eq.\ (\ref{im}) which shapes time
evolution of $R$ and therefore the time dependence of quantum potential.
The initial condition for this equation plays double role. First,
$R^2(x,0)$ is to be interpreted as the probability distribution for
various initial conditions for $x$, and second $R(x,0)$ is itself an
initial condition for $R$ time evolution.

This shows that in the case of this equation we have to do with some
kind of personality split. To cure this disease, we should, in
principle, proceed as follows.

Recall that $\dr{x}{S}=p$. Thus, Eq.\ (\ref{im}) can be rewritten as
$$
\der{t}R+\frac{1}{m}\frac{\dd R}{\dd x} p
+\frac{1}{2m}R\frac{\dd^2 S}{\dd x^2}=0.
$$
If we could express $\frac{\dd^2 S}{\dd x^2}$ as a function of $p$, $q$,
$t$, we would be able to turn the equations governing the quantum
potential dynamics to closed form. Moreover, we will be able manifestly take
into account the back reaction of $p$ and $x$ on the quantum potential itself.
This procedure is currently under investigation, and in this paper I will
ignore
this equation whatsoever assuming that it is identically satisfied.
\newline

To finish this subsection let me make some remarks.

The dynamical equations above describe trajectory of a particle. One may
treat this trajectory as some unphysical (i.e., not related to reality)
auxiliary `picture' of the particle behavior. This point of view is especially
fruitful in the cases when interpretation of the wave function is not clear,
for
example in quantum cosmology \cite{AJ}. However, as stressed by Bohm and Hiley
in
\cite{Bohm}, one may take
the `ontological' stand and assume that the trajectories are real, that is that
there exists a `real' particle moving along the trajectories. It should
be mentioned that even if such an interpretation of quantum mechanics contains
`hidden variables' (the trajectory itself) it does not contradict the Bell
 theorem \cite{Bell}, because the theory is clearly non-local. This fact is
not
by itself very much fearsome since, as  careful analysis of Einstein --
Podolsky -- Rosen
paradox shows, in the standard Bohr - Von Neumann interpretation
of quantum mechanics, the theory
is non-local as well.
\newline

Let me now turn to  more  complex example, which I will call

\subsection{Cosmo quantum mechanics}
In this subsection I will consider an example of the so-called one dimensional
parametric
systems.  Systems like that possess a symmetry which  makes them independent
of a parametrization of the world line; this symmetry is the one dimensional
diffeomorphism invariance which is analogous to the diffeomorphism
invariance in four dimensions. On the classical
level the whole dynamics of the parametric particle is given by one constraint
being the hamiltonian; in quantum case, according to the Dirac method of
quantization of constrained systems, the hamiltonian operator must annihilate
physical state, and thus the wave function is, by virtue of the Schr\"odinger
equation, time independent.

Now consider a parametric model with  simple hamiltonian (which nevertheless
is of importance in quantum cosmology, see \cite{AJ})\footnote{In what
follows all quantum mechanical operators will be written in sans serif
type face: $\sfa$, $\sfb$, \ldots, $\sfA$, $\sfB$ etc.}
\be
\sfH\,\J(x^i) = \lp \half g^{ij}\nabla_i\nabla_j - V(x^i)\rp\,\J(x^i)
=
\lp\half\Box - V(x^i)\rp\, \J(x^i) =
0,\label{ham}
\ee
where $g^{ij}$ may be $x$-dependent.
Let us assume again that $\J$ has the
following
polar decomposition
\be
\J = R(x^i) \mbox{exp}\lp\frac{i}{\hbar} S(x^i)\rp\label{wf}
\ee
with both $R$ and $S$ real. Inserting (\ref{wf}) into (\ref{ham}), we
obtain two equations corresponding to real and imaginary part,
respectively. These equations read
\be
\ch[S(x)] = \frac{1}{2} g_{ij}\dr{x^i}{S}\dr{x^j}{S} + V(x^i) =
\frac{\hbar^2}{2} \frac{1}{R}\Box R,\label{Re}
\ee
\be
R\Box S + 2 g^{ij}\dr{x^i}{S}\dr{x^j}{R} =0.\label{Im}
\ee
Equation (\ref{Im}) will not concern us anymore. As in the previous example
it corresponds to probability conservation. We
assume
that the wave function $\J$ is  a solution of equation (\ref{ham}),
and thus this equation is
identically satisfied (even though we may not know what the explicit
form
of the wavefunction is.) On the other hand, equation (\ref{Re}) is of
crucial importance. This equation can be used to derive the time
dependence and then serves as the evolutionary equation in the
formalism.

For, let us introduce time $t$ through the following equation
\be
\frac{d x^i}{d t} = g^{ij}\frac{\d \ch[S(x)]}{\d (\partial S
/\partial x^j)}.\label{time}
\ee
This equation defines the trajectory $x^i(t)$ in terms of the phase
of the
wavefunction $S$. Now we can substitute back equation (\ref{time}) to
(\ref{Re}). Assuming that the matrix $g^{ij}$ has the inverse
$g_{ij}$,
we find ($\dot x^i = \frac{d x^i}{d t}$)
\be
\half g_{ij}\dot{x^i}\dot{x^j} + V(x^i) = \frac{\hbar^2}{2}
\frac{1}{R}\Box
R.\label{evol}
\ee
We see therefore that the quantum evolution differs from the classical
one only by the presence of the quantum potential term
$$
-V_{quant}(x^i) = \frac{\hbar^2}{2} \frac{1}{R}\Box
R
$$
on the right hand side of equation of motion. Since we assume that the
wave function is known, the quantum potential term is known as well.

Equation (\ref{evol}) is not in the form which is convenient for our
further investigations. To obtain the desired form, we define classical
momenta
$$
p_i =         \frac{\d \ch[S(x)]}{\d (\partial S
/\partial x^i)} = g_{ij}\dot x^j\label{mom}
$$
and cast equation (\ref{evol}) to the form
\be
\ch \equiv \half g^{ij}p_ip_j + V(x^i) - \frac{\hbar^2}{2}
\frac{1}{R}\Box
R = 0.\label{constr}
\ee

We regard $\ch$ as the generator of dynamics acting through the
Hamilton
equations
\bq
\dot p_i &=& - \dr{x^i}{\ch} \nonumber\\
\dot x^i &=& \dr{p_i}{\ch}.                    \label{Hameq}
\eq

The time evolution
is therefore governed by equations (\ref{Hameq}) subject to the constraint for
initial conditions (\ref{constr}). This completes the technical part.
However a number of remarks is in order.

\begin{enumerate}
\item The quantum potential interpretation may be used to obtain a
well
defined semi-classical approximation to quantum theory. Indeed, it
can be
said that the system enters the (semi-) classical regime if the
quantum
potential is much smaller than the regular potential term.
\item One of the major advantages of the quantum potential approach is
that it provides one with an affective and simple way of introducing
time even if the system under consideration has a hamiltonian as one
of
the constraints. In particular this approach serves as a possible
route
to final understanding of the problem of time in quantum gravity.
\item Related to this is the problem of interpretation of wave
functions which are real. This problem has been a subject of numerous
investigations, but from the point of view of quantum potential the
resolution of it is quite simple. We just say that real wavefunctions
(of
the universe) represent a model without time evolution (and therefore
time) at all.\footnote{Recall that the standard (time-dependent) Schr\"odinger
equation does not
have any real solutions.} It is clear from the formalism: $\dot x$ is
just equal to
zero, so nothing evolves and therefore there is no clock to measure
time. On the other hand, equation (\ref{evol}) means that for the real
wave function the system settles down to the configuration for which
the
total potential (i.e., classical plus quantum) is equal to zero.

But there is one important modification of the theory for real wave functions.
Such wave functions, which form a degenerate subclass of all wave functions
being solutions of the theory, should be regarded as ones which impose
additional
constraints on the quantum potential theory, namely that the momentum is zero.
This constraint, together with $\ch=0$ form a second class system. This means,
first, that one can take them as strong equalities (up to the standard
manipulations with Dirac bracket), and this leads to the abovementioned
equality of classical and quantum potential. What is more important, however,
since the hamiltonian is now a second class constraint, it does not generate
gauge transformations (time reparametrization) any more. Roughly speaking, the
quantum potential theory is "anomalous" for real wave functions. It should be
mentioned
that similar effect turns out to be present in the full quantum gravity theory,
and without the interpretation given above it may lead to apparent paradoxes,
one of whose was presented in \cite{Shtanov}. I will discuss the quantum
gravity
case below.
\item The definition of time by equation (\ref{time}) is, of course,
not
unique. In fact, we can  use a more general expression
\be
\dot x^i = N(t) \frac{\d \ch[S(x)]}{\d (\partial S
/\partial x^i)} ,
\ee
where, in the case of gravity, the function $N$ is to be identified
with
the lapse function of ADM formalism.
\end{enumerate}

\section{Wheeler -- De Witt equation and its formal solution}
 Now we can turn to the real thing -- the quantum theory of gravity.
This theory is defined by two sets of constraints, one of whose are generators
of three dimensional diffeomorphisms operators:
\be
{\sfH}_a = - 2 \nabla_b \sfp^b_a ,\lbl{diff}
\ee
and the second consists of a single but very complicated operator, called the
Wheeler -- De Witt operator
\be
{\sfH}_{WDW} = {1 \over 2 \mu} {\cal G}_{abcd} \sfp^{ab}
\sfp^{cd} + \mu \sqrt h \left(2 \Lambda - {}^{(3)} {\cal R} \right). \lbl{WDW1}
\ee
Here
$\mu = \left( 16 \pi G \right)^{- 1}$, $G$ is the Newton's constant,
\be
{\cal G}_{abcd} = {1 \over \sqrt h}
\left(h_{ac} h_{bd} + h_{ad} h_{bc} - h_{ab} h_{cd} \right), \label{wdwmet}
\ee
$\sfp^{ab}$ are momentum operators related to the three metric $h_{ab}$,
${}^{(3)} {\cal R}$ is the three dimensional curvature scalar, and $\L$ the
cosmological constant.

As it stands, this operator is meaningless. It contains the product of
two functional derivatives acting at the same point; as a result,
while acting on a wave function, the
product of two delta functions at the same point appears. This is, of
course  standard property of second order operators in Schr\"odinger
representation. The way out of this problem is to regularize the kinetic
term of the operator, that is, to replace $\sfH_{WDW}$ above by
$$
\sfH_{WDW}^{reg}(x;t) =\lim_{x\rar x'}\left\{ \frac{1}{2\m}\ck_{abcd}(x,x';t)
\sfp^{ab}(x)\sfp^{cd}(y)\right\} +
\m\sqrt{h}\lp2\L-{}^{(3)}\car\rp =
$$
\be
= \frac{1}{2\m}\triangle_{reg}
+ \m\sqrt{h}\lp2\L-{}^{(3)}\car\rp,\lbl{WDWreg}
\ee
where in the limit when the parameter $t$ goes to zero,
\be
\lim_{t\rar0}\ck_{abcd}(x,x';t) = \cg_{abcd}(x)\d(x-x').\lbl{ini}
\ee
The function $\ck$ can be found, for example, by using the heat kernel
equation
\be
\der{t}\ck(x,x';t) = \nabla^2_{(x')}\ck(x,x';t)
\ee
with the initial condition (\ref{ini}). This approach was proposed
in \cite{mansfield} in the context of
Yang -- Mills theory, and then applied to the theory of gravity in
\cite{3jap}. In this latter paper the action of the regularized WDW
operator on some simple expressions built from the metric and curvature
was also found. The virtue of this approach is that mathematically
meaningless expressions (like products of delta functions at the same
point) are now replaced by well controllable singularities of $\ck(x;t)=
\lim_{x\rar x'}\ck(x,x';t)$ for
small $t$. Each singular term can be then replaced by a renormalization
constant. The problem is however that at the first glance it seems that
the number of such independent constants is infinite. However, the
action of the operator has been tested on some simple expressions (see
\cite{3jap}) and one may hope that the physical wave functions, i.e.,
the wavefunctions annihilated by all constraints (or, to be more precise,
the matrix elements constructed with the help of them) will
eventually contain only finite number of them (hepefully two -- one
corresponding to the renormalization of the Newton's constant, and
second to the renormalization of the cosmological constant.)\footnote{I
would like to thank Jeff Greensite for long discussion concerning this
point.}

Be as it may, let us assume that we have constructed the regularized
Wheeler -- De Witt operator. Then one can easily write down the formal
expression for the physical wave function being annihilated by this
operator. This can be done by observing that a physical state is of
the form\footnote{The commutator of two WDW operators is proportional to
the diffeomorphism operators (see \cite{3jap} for discussion of the
algebra of regularized constraints in quantum gravity.) Therefore it is
not needed to apply diffeomorphism constraints separately.}
\be
\J_{phys}=\d\lp\sfH_{WDW}^{reg}\rp \j[h]
,
\ee
where $\j[h]$ is any wave function(al) of the metric, that is
\be
\J_{phys}= \int \cd N(x)\; e^{i\int\, d^3x
N(x)\sfH_{WDW}^{reg}(x)}\;\j[h].\lbl{wfpi}
\ee
Even though the expression above is very formal, it can be used to obtain a
perturbative expansion (in $\m$) of a solution. Second, and this will be
quite important in our analysis below, using this formula we can
find out some properties of the physical wave function.

The next important property of the formula (\ref{wfpi}) is that it may be
used as the starting for construction of the inner product. It should be
recalled that one of the major problem of the Dirac quantization
procedure is that it does not offer any constructive way of finding what
the inner product of physical states is. I will not dwell on this
problem any longer as it is the subject of the forthcoming paper
\cite{jurinner}. Let me only mention that some partial results and
discussion are contained in \cite{AJ}.
\newline

Now we are ready to extend the discussion of one dimensional parametric
systems presented in subsection 3.1 to the case of full theory of
gravity. This will by the subject of the next section.

\section{Quantum potential for quantum gravity}

As in subsection 3.1 we start with decomposition of the wave function to
the polar form (from now on $\hbar=G=c=1$)
$$
\J[h] = R[h]e^{i S[h]}.
$$
Substituting this into the WDW equation, one easily finds (see also
\cite{Shtanov})
\be
-\frac{1}{2\m}\ck_{abcd}\vder{S}{ab}\vder{S}{cd} +
\m\sqrt{h}\lp2\L-{}^{(3)}\car\rp +
\frac{1}{2\m}\frac{\triangle_{reg}\, R}{R} =0.
\ee
If one identifies momenta with the (functional) gradient of $S$, to wit
\be
p^{ab}(x) = \vder{S}{ab}(x),
\ee
this equation turns to the Hamilton--Jacobi equation for general relativity
\be
-\frac{1}{2\m}\cg_{abcd}p^{ab}p^{cd} +
\m\sqrt{h}\lp2\L-{}^{(3)}\car\rp +
\frac{1}{2\m}\frac{\triangle_{reg}\, R}{R} =0,\lbl{HJ}
\ee
with the additional last term corresponding to quantum potential.
This equation (without the potential term, of course) was first analyzed by
Gerlach \cite{gerlach}.
Equation (\ref{HJ}) is only one of two equations which
result from action of the WDW operator on the wave function. As in the
case of the parametric particle mechanics, we ignore the second
equation, which could be interpreted as an equation shaping the quantum
potential and/or guaranteeing conservation of probability.

The wave function is subject to the second set of equations, namely the
ones enforcing the three dimensional diffeomorphism invariance. These
equations read (after decomposing into real and imaginary part)
\be
\nabla^a\vder{S}{ab} = \nabla^a\, p_{ab} =0\lbl{3diff}
\ee
\be
\nabla_a\vder{R}{ab} = 0
\ee
Each of these equations has different interpretation. The second
expresses invariance of $R$ with respect to spatial diffeomorphisms --
this invariance is actually guaranteed by the fact that $R$ is the
modulus of a wave function being a solution o WDW equation; the
first is to be interpreted as a constraint equation.

Thus our theory is defined by two equations (\ref{HJ}) and
(\ref{3diff}). Now we can follow without any alternations the derivation
of Gerlach \cite{gerlach} to obtain the full set of ten equations
governing the quantum gravity theory in quantum potential approach
\bq
0= \ch^a &=& \nabla_a\, p^{ab} ,\lbl{diff1}\\
0= \ch_{\bot} &=&
-\frac{1}{2\m}\cg_{abcd}p^{ab}p^{cd} +
\m\sqrt{h}\lp2\L-{}^{(3)}\car\rp +
\frac{1}{2\m}\frac{\triangle_{reg}\, R}{R},\lbl{HJ1}\\
\dot{h}_{ab}(x,t) &=& \left\{ h_{ab}(x,t),\, \ch[N,\vec{N}]\right\},\lbl{1}\\
\dot{p}^{ab}(x,t) &=& \left\{ p^{ab}(x,t),\, \ch[N,\vec{N}]\right\}.\lbl{2}
\eq
In equations above, $\{\star,\, \star\}$ is the usual Poisson bracket, and
\be
\ch[N,\vec{N}]=\int\, d^3x \lp N(x)\ch_{\bot}(x) + N^a(x)\ch_a(x)\rp
\ee
is the total hamiltonian (which is a combination of constraints).
\newline

A number of comments is in order
\begin{enumerate}
\item Let us observe that we somehow "solved" the problem of time in
quantum general relativity. Indeed out of sudden time appears in Eqs.\
(\ref{1}), (\ref{2}). It is a matter of taste if this is to be
considered as a solution of this problem, nevertheless, in this
formulation the problem of time in quantum general relativity
is as simple (or as difficult) as the analogous problem in classical
theory of gravitation (see \cite{Isham2}.)
\item The very important problem related to Eqs.\ (\ref{diff1} --
\ref{2}) is under which conditions these equations are equivalent to ten
four dimensional covariant equations ("quantum Einstein equations".) To
see what this problem is about, let us observe that Eqs.\ (\ref{diff1}),
(\ref{HJ1}) are constraint
equations. This means that $\ch_a$ and $\ch_{\bot}$ must have (weakly)
vanishing Poisson bracket with $\ch[N,\vec{N}]$. But this means that
Poisson brackets of $\ch_a$ and $\ch_{\bot}$ must form closed algebra.
Yet it is well known that it is quite difficult to close the bracket
$$
\left\{ \ch_{\bot}(x),\, \ch_{\bot}(y)\right\}
$$
for arbitrary quantum potential. Indeed one can easily check that this Poisson
bracket does not close if $\ch_{\bot}$ contains, for example, terms quadratic
in
curvature.\footnote{This fact can be understood by observing that the effective
theory in four dimensions is, in this case, a higher derivative theory.} Now,
a simple inspection of Eq.\ (23) shows that $\J_{phys}$ {\em will} contain
higher
order terms.
It should be observed however that
quantum potential term $\frac{1}{2\m}\frac{\triangle_{reg}\, R}{R}$ is
essentially non local and the standard argument may be not applicable.
This problem will be the subject of the separate paper.

Assuming however that for some quantum potentials the bracket above does
not close. Then we have to do with secondary constraints which will be
second class (this follows from simple counting of degrees of freedom.)
This just means that the $3+1$ symmetry is not promoted to the full four
dimensional diffeomorphism invariance after quantization. This fact is without
doubts very
important, for example it may be related to the appearance of minimal
length in quantum gravity (see \cite{garay}). It may also mean that
either $3 + 1$ splitting being the first step of quantization procedure is
incorrect and/or that the four dimensional diffeomorphism invariance is
just a low-energy phenomenon.
\item Talking anomalies, let us return to the apparent puzzle discussed
in \cite{Shtanov}. In this paper the author argues that for real solution
of WDW equation there exists the transformation generated by the hamiltonian
$\ch_{\bot}$ which transforms the solution of Eqs.\ (\ref{1}), (\ref{2}) into
configuration which is not a solution.
But if the wave function we start with is real, the
momentum of gravitational field vanishes. Then the quantum potential
must compensate the classical "potential term" -- the three-curvature.
As in the case of parametric particle, this means that the constraint
$\ch_{\bot}$ together with $p^{ab}$ becomes a second class system and
the time shift invariance of the theory is lost.
\item Let me stress once again that one of the crucial points of the
construction is an appropriate regularization of the kinetic
(derivative) term in $\ch_{\bot}$. This problem is not
discussed in most papers concerning (semiclassical) quantum gravity in
metric representation. However without proper understanding of
regularization and renormalization of the theory it is impossible to
find any solution and to even start the program described above.
\item Finally, it is feasible to extend the above discussion to the case
of Ashtekar variable (in either connection or tetrad representation.)
This is especially important because one clearly should include matter
fields and it is well known that we need tetrad formalism to take care
of fermions (for bosonic matter fields there are only obvious technical
modifications of the formalism.)
\end{enumerate}

To conclude, the qunatum potential program may provide us with an important
insight into the meaning of the physical wave function of quantum gravity.
However, we need to find a wave function of the universe first. As for now,
therefore, this set of ideas can be only applied in the context of the
minisuperspace models. Let us hope that this situation will soon change.

\section{Acknowledgement}
I would like to thank Prof.\ Jeff Greensite and
Mr.\ Arkadiusz B\sll{}aut for discussions concerning the problems discussed
in this paper.

\section{Bibliography}

\newcommand{\prev}[3]{Phys.\ Rev.\ {\bf {#1}},  {#2}, ({#3})}
\newcommand{\prep}[3]{Phys.\ Rep.\ {\bf {#1}},  {#2}, ({#3})}
\newcommand{\prevD}[3]{Phys.\ Rev.\  {\bf D {#1}},  {#2}, ({#3})}
\newcommand{\pletB}[3]{Phys.\ Lett.\  {\bf  {#1} B},  {#2}, ({#3})}
\newcommand{\nphB}[3]{Nucl.\ Phys.\  {\bf B {#1}},  {#2}, ({#3})}

\end{document}